# Inland waterway transport accident analysis of Bangladesh: based on location, time, and regression approach.

Ahammad Abdullah[1], Md. Jubair Mia[2*],

[1]PhD student, Florida Atlantic University, USA
[2]Department of Naval Architecture and Offshore Engineering, Bangabandhu Sheikh Mujibur Rahman Maritime University, Bangladesh

## ABSTRACT

Bangladesh, situated in the foothills of the Himalayas in South Asia, is a nation characterized by its extensive river network. This riverine state comprises various features such as small hill ranges, meandering seasonal creeks, muddy canals, picturesque rivers, their tributaries, and branching streams. Numerous cities and ports have been established along both sides of these rivers, forming an inseparable connection to the country's civilization and agricultural system. These waterways serve as vital channels for communication and transportation of goods due to their easy accessibility and cost-effectiveness. In Bangladesh, waterway accidents have been a persistent issue. Every year, a significant number of people suffer injuries, fatalities, or go missing due to shipping accidents. While major naval accidents briefly generate public outcry and prompt investigations by the media and authorities, this study aims to examine accident data spanning from 1995 to 2019. The goal is to identify the primary factors contributing to these accidents using multiple linear regression theory in various mathematical combinations. The model's accuracy is validated using different datasets. Additionally, this study delves into the identification of accident-prone areas and the temporal distribution of accidents, offering a comprehensive understanding of accident occurrences. The findings of this research will aid stakeholders and authorities in making informed decisions to prevent waterway accidents. Furthermore, the study presents recommendations focused on accident path planning and avoidance, as well as establishing a theoretical foundation for driver assistance systems.

## 1. INTRODUCTION

In this ancient world of 400 million years, the age of Bangladesh is at most 3 to 4 million years. In other words, Bangladesh is at the level of the world's youngest children. About 700 rivers with branches and branches flow through the country occupying an area of 24,140 km with huge amount of water. They cover nearly 7% of the total area of Bangladesh. Bangladesh has about 9000 square kilometers of territorial waters with a 7200 long coastline (Banglapedia, 2021). The length of navigable inland waterways is approximately 5,968 km, which decreases to 3,600 km during dry seasons (BIWTA, 2021). Inland waterways have been categorized into four categories based on their traffic and economic value (BIWTA, 2021).

According to long-term technology 'Delta Plan-2100' developed by the Government of Bangladesh, river transport system is an integral pathway for economic growth and development. Besides, the waterway transportation network is both extensive and well-connected with the rest of the transportation system. People are becoming more dependent on the river transport system. Almost 50,000 people use inland waterways for traveling and transporting merchandise, especially in the country's southern districts. More than 50% of the total merchandise and one-fourth of the total passenger traffic in Bangladesh is carried by inland water transport (BIWTA, 2021)

There are three departments of the government namely Ministry of Shipping (MOS), Department of Shipping (DOS) and Bangladesh Inland Water Transport Authority (BIWTA) which are directly or indirectly concerned with the safety of sea voyages. The Ministry of Shipping is entrusted with the responsibility of formulating policies related to the development and management of waterways. And the Department of Shipping (DOS) is engaged in the responsibility of checking the fitness of the vehicles with the power of the government to ensure the safety of all types of vessels. BIWTA on the other hand is engaged in ensuring the safety and meteorological information transmission of navigational and vehicle anchorage-related sites. It is also reported that due to shortage of manpower and necessary modern equipment, the expected progress of work in the departments is being hampered. According to BIWTA, the number of registered boats in the country is only 12,959. Apart from this, many unregistered boats are regularly plying the waterways and many accidents are also caused by them.

According to Akhter (2017), almost 5407 people died in 600 casualties during the last 40 years. But the death toll is not decreasing; rather, it is increasing devastatingly. In recent years, the death rate is much higher than the previous years. However, to ensure smooth and safe navigation there is the 'Inland Shipping Ordinance-1976', which lays down all matters relating to shipping or transport laws and the provision of punishment for violators. None of the drivers, vehicle owners or passengers is aware of this, so we have to see the tragic scene of accidents again and again due to arbitrary movement of vehicle owners, drivers and passengers. There are so many reasons behind the accidents, such as carrying cargo or passengers in excess of capacity, negligence and inefficiency of drivers, defects in parts, hostile weather, defective structure of vehicle, unsafe sea route, expired fitness of vehicle, collision between ships, unlicensed operator, inadequacy of radar or radio equipment, excessive current, deficiencies in monitoring or control, inapplicability of exemplary punishment, lack of monitoring, overloading, natural disasters, lack of awareness, etc.

This paper aims to investigate the causes behind the inland waterway accidents of Bangladesh through different analysis methods like analysis based on location, analysis based on time, and a new approach to predict the number of accidents by regression analysis. This research also discusses the most effective causes (independent variables) of the accident, which increase the response variable (number of accidents). After finding the root of the accident, some propositions will be given to the maritime authority to ensure the inland waterway safety of Bangladesh through reduce the number of casualties.

## 2. LITERATURE REVIEW

According to the information provided by the Directorate of Maritime Transport of the Government of Bangladesh, 608 boat accidents occurred in Bangladesh from 1991 to 2018 in which 3 thousand people died. Among them, there are 1,781 launch passengers.[1] And according to the report of the National Committee for the Protection of Sea, Road and Railway, a private organization, there were 653 small and big launch accidents in the 20 years from 1996 to 2015, in which the number of casualties was 6,408.[2] On the other hand, the river According to the safety social organization Nogar, in the 14 years from 2005 to 2019, more than 6,000 people lost their lives in 535 major marine accidents and 863 committees were formed to investigate these accidents.[3] Very few reports of these investigation committees have seen the light of day. During this research work, several studies have been investigated and it has been found that most studies highlighted water transport accidents as isolated events by blaming authorities related to maritime safety rather than describing the problem as a system error. These studies are deficient in terms of detailed analysis of accidents based on various causal factors.

Khalil and Tarafder (2004) studied ferry diasters in the rivers of Bangladesh and suggested some modifications in vessels design for the acquisition of additional initial stability due to an upward change in the center of gravity to keep the vessel safe from capsizing.

Baten (2005) discussed irregularities in the maritime transportation sector from safety viewpoint and criticized the current rules of Inland Shipping Ordinance (ISO 1976) for being inept, and recommended certain reforms of the rules.

Chowdhury (2005) devolved a GIS-based accident information system for analyzing maritime accidents. Huq and Dewan (2006) suggested to ensure radar and radio equipment's in all types of passenger vessels. The study also recommended to establish efficient marine court to ensure the passenger safety.

Awal et al (2007) found that relatively higher percentage of maritime accidents occur during the monsoon period due to adverse weather. The study also showed that accident rate is slightly higher during business hours. The study recommended to improve the weather forecasting system of Bangladesh.

Iqbal et al. (2008) proposed to ensure additional above water reserve buoyancy in small vessels. The study suggested modifying the weather criterion considering wind speed, rolling of the vessels, and crowding of people on one particular side of the vessel.

A study by Azad (2009) found the primary reasons behind maritime accidents are-recruitment of unskilled masters and crews, relevant government bodies are understaffed, and maintenance of the vessels are low quality.

Rahman and Rosli (2014) developed a model to detect overloading in inland vessels by using elevator concept. The author also discussed the necessity of implementing overloading detection systems in inland vessels.

Uddin et al (2017) analyzed inland waterway accidents that occurred from 2005 to 2015 and found that almost two-thirds of the accidents happened due to collisions between two vessels and 45% of the vessels sunk after the accidents. The study is recommended to ensure proper investigation of accidents and to create mass awareness among the people.

Another study by Uddin et al (2017) found that cargo vessels tend to face more accidents than other types of inland vessels. The study also showed that fatality compromised a relatively higher portion of casualties from 2005 to 2015.

Rahman (2017) analyzed maritime accidents that occurred from 1981 to 2015 and found that the rate of accidents has increased dramatically in recent years. The study proposed that every vessel must have sufficient navigational visibility to avoid collision during foggy weather. The study also suggested that all passenger vessels should have a common structural design approved by the department of shipping.

Probha (2017) conducted a comparative analysis of passenger safety in highways, railways, and waterways and found that waterway accidents contributed a significant percentage (about 37%) of the total accidents from 2008 to 2015.

Mia et al. (2021) analyzed the characteristics of maritime accidents that occurred during the last 25 years and found that collision is the main reason for waterway accidents in Bangladesh. This study also performed a distinctive analysis of maritime accidents that occurred from 2010 to 2019 and showed that the rate of accidents has increased by 7.5% in recent years but the death rate has decreased by 23.2%.

A popular model for explaining the causality of various accidents is the "Swiss cheese model". Later in the 1980s and 1990s, the "Tripod Beta" model developed by the joint research of the University of Leiden in the Netherlands and the University of Victoria in the United Kingdom thinks that there are various preconditions behind the wrong behavior of people that act as an influence behind the accident. Although its results in accident prevention are far-reaching; solving managerial, process and organizational cultural problems often takes a lot of time and money.[4] And so in this research, the authors have tried to construct a new methodology to explain the causes of inland waterway accidents in Bangladesh and this methodology or procedure is totally different from the above-discussed way of analysis of accident research. The accident causes will be explained by the mathematical model as well as the intensity of the causes also. In this regard, different parameters are considered as independent variables including number of collision, stormy weather, excessive current, etc and the outcome variable is number of accidents. It is called also response variable. The process developed in the work can be used to formulate full-scale mathematical model to predict the number of accidents.

## 3. DATA COLLECTION

The main purpose to perform this research is to find out the true factor behind all those maritime accidents that occurred in Bangladesh in the last 25 years through regression analysis. While conducting the research, the most difficult and brain hammering job was to collect authentic data and organize them in a suitable manner. Moreover, the missing and lack of detailed information made this job extremely tough. We gave our top priority to collect the data from authentic sources. Our primary sources were the Department of Shipping (DOS), Bangladesh Inland Water Transport Authority (BIWTA). Again, we collected data from the Accident Research Institute (ARI) of Bangladesh University of Engineering Technology (BUET) and some renowned daily newspapers of Bangladesh. It was observed that DOS and BIWTA collect data only for legal purposes, so it was never easy to prepare the data for scientific uses. Daily newspapers emphasize over subjective matters, technical parameters are usually not a part of their report. So, creating the database was so time consuming. However, we needed an authentic database to justify ours. Mia et al. (2020) created a database that is similar to that is developed for this study. We gave an eye over the accident location and time to understand the matter in-depth, and we considered the primary reasons of accidents as variable for regression analysis. Taking into account all, we categorized the reasons into six groups. We considered every maritime accident that occurred between 1995 to 2019. The regression analysis has been performed by Microsoft Excel software and graphical forms were illustrated by Matlab and ArcGIS. Attempts have been made to keep the representation of the results as clear and unambiguous as possible.

## 4. METHODOLOGY

Using the collected data, this research is conducted in three different ways which can analyze the causes of accidents very precisely from the discussed earlier method. **Firstly,** analysis is done on location based and is a very much effective way to identify the stretch point by sorting the highest number of accidents. **Secondly**, the analysis is performed based on time of the accident. In this way, a specific time is found when the maximum number of accidents has occurred. **Finally**, regression analysis is applied to predict the number of accidents due to the different causes which are closely related to the occurrence of the accident. In the regression analysis predicted the number of accidents is considered as the dependent variable and the causes of accidents are considered as independent variables.

### 4.1 DISTRICT WISE DISTRIBUTION OF ACCIDENTS

By analyzing the waterway routes and accidents, very alarming results are found. It is seen that the number of accidents has a proportional relation with the density of vessels plying on a particular river route. Barishal and Chattogram are the busiest water transportation routes for Bangladesh. Unfortunately, these two routes faced the maximum number of accidents during the last 2 decades. After that, Dhaka, Naryanganj, Chandpur, Khulna, and Munshiganj are the most accident-prone districts in Bangladesh. A total of 26 districts are found to face at least one water transport accident in the last 25 years, which covers almost every important riverways of Bangladesh.

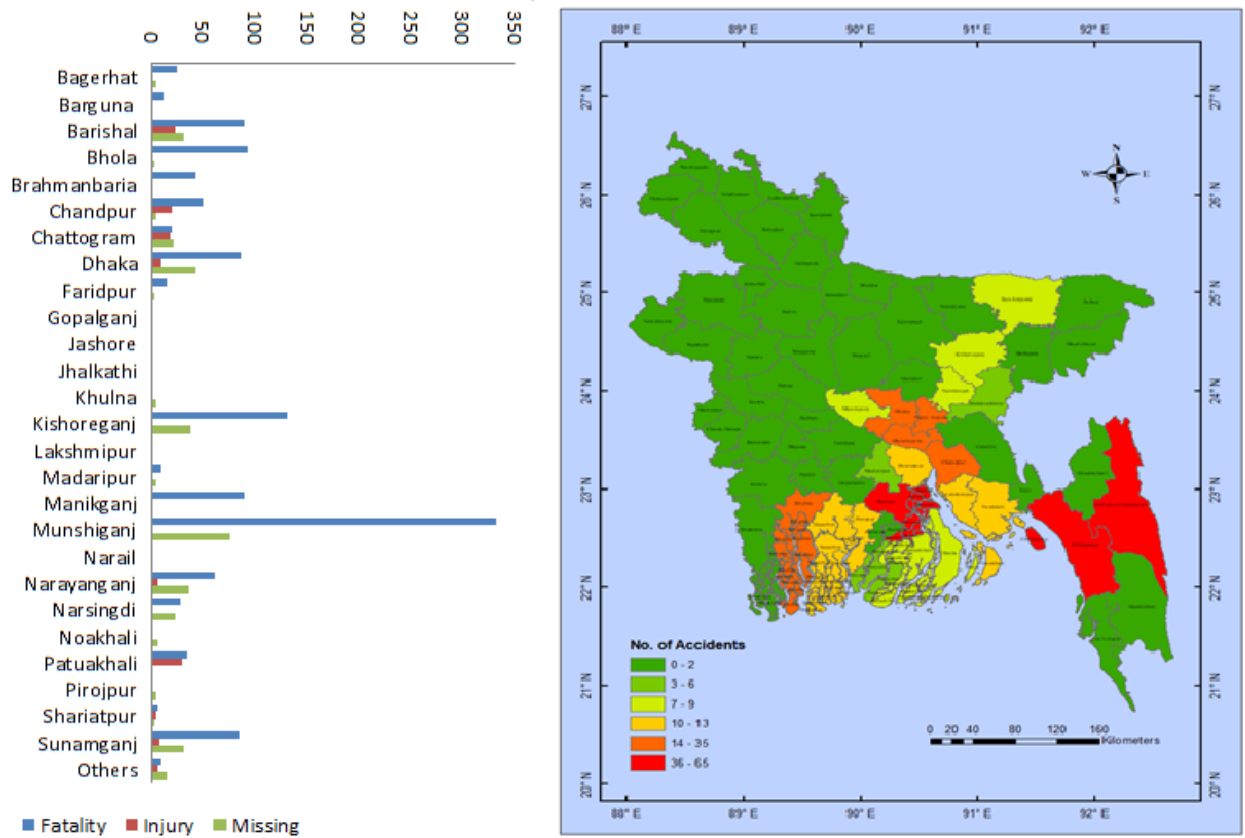

**Figure 1: Location wise distribution of waterway accidents.**

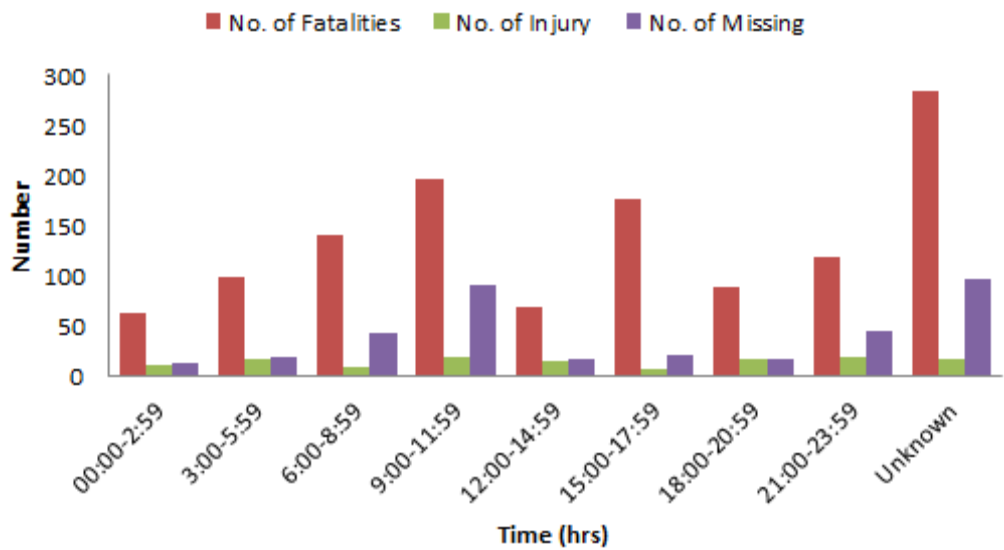

**Figure 2: Timewise distribution of waterway accidents.**

## 4.2 HOURLY DISTRIBUTION OF ACCIDENTS

Two-thirds of the accidents that occur on Bangladesh's waterways are due to collisions between two vessels. Most of these collisions occur at night or in fog during winter when visibility is limited.[5] The hourly distribution of accidents is shown in the figure 2. The figure indicates that accidents frequently occur all around the day. It also illustrates that during the Post Meridiem (PM) period, more accidents are occurring than during the Ante Meridiem

(AM) period. It is seen that the maximum number of accidents took place between 10 AM to 4 PM, which is very alarming. Because this period is the business hour in Bangladesh and thousands of passengers and business, people use waterways during this period. It is also observed that a noteworthy portion of accidents occurred from 6 PM to midnight. Because the majority of Bangladeshi vessels have lacking's of adequate navigational and lighting facilities. Newspapers often report on launches being stuck underwater, collisions between stuck launches and other launches, lack of sufficient beacon lights on waterways. The launch should have radar for direction finding, GPS, echo sounder for measuring the depth of the river bed, fog lights for fog, high power VHF radio for ship to ship communication, electric wheel for speed control of the ship etc. But even though there are 695 passenger ships running across the country, half of them do not have these equipments. Out of 70 luxurious launches, only 10 launches are suitable for navigation in thick fog.[6] Therefore the vessels plying at night are highly exposed to the risk of accidents. Though the occurrence time of the significant number of accidents is unknown. The reason is very clear because most of the accidents are not investigated on time in Bangladesh, and there is lacking proper documentation.

**4.3 REGRESSION ANALYSIS TO DEVELOP MATHEMATICAL MODEL**

More than one independent variable is required in the regression model in most research subject matter when regression analysis is used. The complexity of most scientific procedures necessitates the use of a multiple regression model to anticipate a critical response.[7] Using the least-square technique, this study seeks to build a multiple linear regression mathematical model. The mathematical model has been developed to determine the predicted number of accident as well as the most prominent cause of inland waterway accidents through regression analysis. This section discusses the statistical relationship between the number of inland waterway accidents (as dependent variable) and the prominent causes of accidents (as independent variables). For the development of mathematical model a linear equation is chosen which is given below.

$$IWA_N = f(A_C, A_{SW}, A_{EC}, A_G, A_O)\dots\dots\dots\dots\dots\dots\dots\dots\dots\dots\dots\dots\dots(1)$$

Where $IWA_N$ represents the number of inland waterway accidents which is the dependent variable. $A_C, A_{SW}, A_{EC}, A_G, A_O$ Indicate the number of accidents that occurred due to collision, stormy weather, excessive current, grounding, and overloading respectively and these causes are the independent variables.

A multiple linear regression model is one in which the coefficients are linear. The multiple linear regression approach sets up the relationship between the dependent and independent variables in order to predict the dependent variable against a collection of independent variables using data from both. The regression model in this study is considered as the following:

$$y = f(x_1, x_2, x_3 \cdots, x_k)\dots\dots\dots\dots\dots\dots\dots\dots\dots\dots\dots\dots\dots(1)$$

Here, *y* represents the dependent variable that is the number of inland waterway accidents and $x_1, x_2, x_3$..... etc. are independent variables such as accidents due to collision, stormy weather, overloading, etc. There can be *k* number of independent variables.

Equation 1 can be written as follows considering all the variables:

$$y_i = b_0 + b_1 x_{1i} + b_2 x_{2i} + \cdots + b_k x_{ki} \ldots\ldots\ldots\ldots\ldots\ldots\ldots\ldots\ldots\ldots\ldots\ldots\ldots (2)$$

Where, $i = 1, 2, 3\ldots\ldots, n$ and $b_1, b_2, b_3, \ldots, b_k$ are the regression coefficients that need to be determined to construct the multiple linear regression model.

The multiple linear regression model can be expressed with a residual term as:

$$y_i = b_0 + b_1 x_{1i} + b_2 x_{2i} + \cdots + b_k x_{ki} + e_i \ldots\ldots\ldots\ldots\ldots\ldots\ldots\ldots\ldots\ldots\ldots (2)$$

In this case, $e_i$ is the model's residual with the lowest possible value. As a result, the sum of errors (SSE) should be kept minimum. SSE can be expressed as follows:

$$SSE = \sum_{i=1}^{n} e_i^2 = \sum_{i=1}^{n} (y_i - b_0 - b_1 x_{1i} \cdots - b_k x_{ki})^2 \ldots\ldots\ldots\ldots\ldots\ldots\ldots\ldots (4)$$

Different coefficients $b_0, b_1, \ldots, b_k$ are used to differentiate the equation (4). The results are set to zero, and the equations are found as follows. The total number of equations available are $k+1$.

$$n*b_0 + b_1 \sum_{i=1}^{n} x_{1i} + b_2 \sum_{i=1}^{n} x_{2i} + \ldots\ldots\ldots + b_{ki} \sum_{i=1}^{n} x_{ki} = \sum_{i=1}^{n} y_i \ldots\ldots\ldots\ldots\ldots\ldots (5)$$

$$b_0 \sum_{i=1}^{n} x_{1i} + b_1 \sum_{i=1}^{n} x_{1i}^2 + b_2 \sum_{i=1}^{n} x_{1i} x_{2i} + \ldots\ldots\ldots + b_{ki} \sum_{i=1}^{n} x_{1i} x_{ki} = b_{ki} \sum_{i=1}^{n} x_{1i} y_i \ldots\ldots\ldots\ldots (6)$$

$$b_0 \sum_{i=1}^{n} x_{2i} + b_1 \sum_{i=1}^{n} x_{1i} x_{2i} + b_2 \sum_{i=1}^{n} x_{2i}^2 + \ldots\ldots\ldots + b_{ki} \sum_{i=1}^{n} x_{2i} x_{ki} = b_{ki} \sum_{i=1}^{n} x_{2i} y_i \ldots\ldots\ldots\ldots (7)$$

...............................................................................................................................
...............................................................................................................................
...............................................................................................................................

$$b_0 \sum_{i=1}^{n} x_{ki} + b_1 \sum_{i=1}^{n} x_{ki} x_{1i} + b_2 \sum_{i=1}^{n} x_{ki} x_{2i} + \ldots\ldots\ldots + b_{ki} \sum_{i=1}^{n} x_{ki}^2 = b_{ki} \sum_{i=1}^{n} x_{ki} y_i \ldots\ldots\ldots\ldots (8)$$

The values of required statistical parameters can be calculated using the collected sets of data. The above equations can be translated into concurrent equations by using certain values. By solving the concurrent equation using any standard method, the values of regression coefficients and the appropriate predicted model's equation can be found.

Various statistical parameters have been considered in order to demonstrate the adequacy of this mathematical model. Considered criteria and statistical parameters for this study are discussed below in detailed.

**4.3.1 Variance Inflation Factor (VIF)**

Multiple linear regressions presuppose that the data does not have any multicollinearity. When the independent variables are very favorably associated with each other, multicollinearity occurs. When there is a correlation between predictors, the standard error of the predictor coefficients increases, and the variance of the predictor coefficients increases. The VIF is a tool that may be used to calculate and measure how much the deviation has risen. The following formula is used to determine VIF:

$$VIF = \frac{1}{1-R^2}$$

Where $R^2$ is the regression model's coefficient of determination. If the VIF value is 5 (five) or above, the independent variables are significantly associated with one another, then the regression model should be discarded. [8]

**4.3.2 Coefficient of Determination ($R^2$)**

The coefficient of determination, also known as the coefficient of multiple determination, is a measure of how much variation in a dependent variable (healing time) is caused by a group of independent variables (repair works and various other parameters). The coefficient of multiple determination is expressed as follows according to Levine et al. (1997) [9]:

$$R^2 = \frac{SSR}{SST}.$$

Where
 SSR= The regression sum of squares. Theoretically, SSR is equal to the sum of the squared differences between the predicted value of y and, the mean value of y
 SSE= SSE is the sum of the squared differences between the observed value of y and the predicted value of y.
 SST= The total of SSR and SSE

In addition to $R^2$, the adjusted $R^2$ is an important criterion in multiple linear regression. It is used to compare two or more regression models that forecast the same dependent variable. The corrected $R^2$ expression is as follows:

$$R^2(adj) = 1 - \left[ (1-R^2)\frac{n-1}{n-k-1} \right]$$

Where,
 $R^2$ = Coefficient of multiple determination
 $k$ = Number of variable
 $n$ = Number of sample

### 4.3.3 Standard Error:

The standard error is a statistical measure of an estimate's statistical correctness, equivalent to the standard deviation of a large population of such estimates. The following is taken into account in this research:

$$s = \sqrt{\frac{SSE}{n-k-1}}$$

The symbols have the usual meaning as explained above.

### 4.3.4 F-Statistics

The F-statistic is the test statistic for F-tests. In general, an F-statistic is a ratio of two quantities that are expected to be roughly equal under the null hypothesis, which produces an F-statistic of approximately. For passing the model $f$ must be greater than $f_{critical}$.

$$\text{F-statistic}(f) = \frac{\frac{SSR}{k}}{\frac{SSE}{n-k-1}} = \frac{\frac{SSR}{k}}{s^2}$$

### 4.3.5 Mallow's CP Statistic

Mallow's Cp statistic is practiced to evaluate the fit of a regression model through ordinary least squares. It is adopted in the context of model selection for predicting some outcome, and the objective is to determine the best model involving a subset of these predictors. A small value of Cp means that the model is relatively precise. The Cp statistic is figured as[10]:

$$c_p = \frac{SS(RES)_p}{MSE_{all}} + 2(p+1) - n$$

Where,

$p$ = Number of independent variables
$SS(Res)_p$ = residual sum of the square from the p number of the variable subset.
$MSE_{all}$ = Mean square error
n = Number of samples

### 4.4 MODEL DEVELOPMENT STEPS

A method is developed to find the most prominent cause of inland waterway accidents, applying the discussed multiple regression theory. In this study, at first independent variables are selected. These variables have significant role in predicting the number of accident. After that, the multicollinearity of independent variables is checked. This checking is done by Variation Inflation Factor (VIF). Then linearity of the variables is checked by scatter plots which is done between dependent and independent variables. Later stepwise regression is performed to choose the best subset of independent variables. After choosing the best independent variables, the selection of the best fit model is completed on the basis of fulfilling the requirement of all statistical parameters (R- Squared, Adjusted R-squared, MSE, and Cp). After choosing the best subsets, a final analysis is conducted to examine the residual versus predicted variable pattern, and independent variables, which is called residual analysis.

If any violation is found in linearity or other assumptions, then transformation of variables is needed. After that, the model is finalized to predict the number of accident as well as identified influential reasons for inland waterway accidents in Bangladesh. The flowchart of this method is shown in the figure.

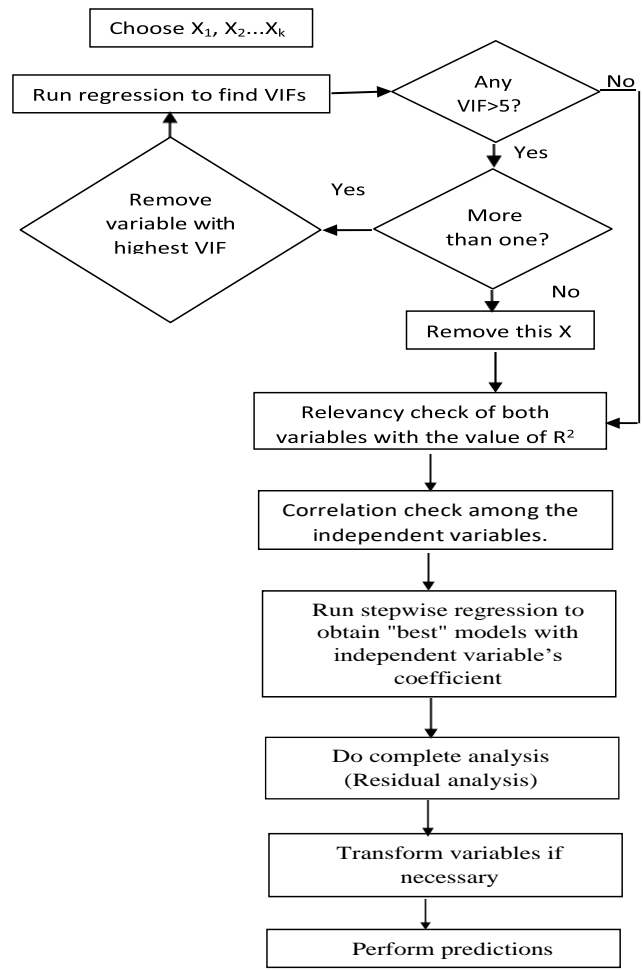

**Figure 3: Model development flow-chart**

According to the model development steps, the collinearity statistics (Variation Inflation Factor, VIF) of the independent variables is shown in table 1. Table 1 shows that no values of VIF cross the limiting value previously mentioned.

**Table 1: Illustration of VIF values for different independent variables.**

| Independent Variable | Collinearity Statistics VIF (Including all variables) |
|---|---|
| Collision (C) | 1.130 |
| Stormy Weather (SW) | 1.159 |
| Excessive Current (EC) | 1.077 |
| Grounding (G) | 1.035 |
| Overloading (O) | 1.165 |

## 4.5 CORRELATION CHECK WITH MULTIPLE R AND SCATTER PLOT

After checking the multicollinearity of all the independent variables in the previous section, correlation check is performed to understand the relation between dependent and independent variables. Correlation value between dependent (Number of accident) and independent variables (Collision, Excessive Current, Stormy, Overload and Grounding) are good that is given below and this value is called Multiple R. From the table 2, it is found that the correlation value of collision, stormy weather and excessive current are good that means these independent variables have strong relationship with dependent variable. Other two independent variables have less strong relation with dependent variable but those two variables pass the criteria of VIF.

**Table 2: Summary of relevancy checks between dependent and independent variables.**

| Dependent Versus Independent Variables | | | Multiple R |
|---|---|---|---|
| Number of accident | vs | Excessive Current | 0.36565 |
| Number of accident | vs | Overload | 0.117473 |
| Number of accident | vs | Grounding | 0.109545 |
| Number of accident | vs | Stormy Weather | 0.314643 |
| Number of accident | vs | Collision | 0.558 |

Figure 4 shows the scatter plot of the dependent variable (number of accidents) versus independent variables. From the figure, it is shown that all the independent variables have a positive relationship with the dependent variable having a positive slope. The value of Multiple R and the positive slope from scatter plot represent that all the independent variables are fit for the multiple regression analysis.

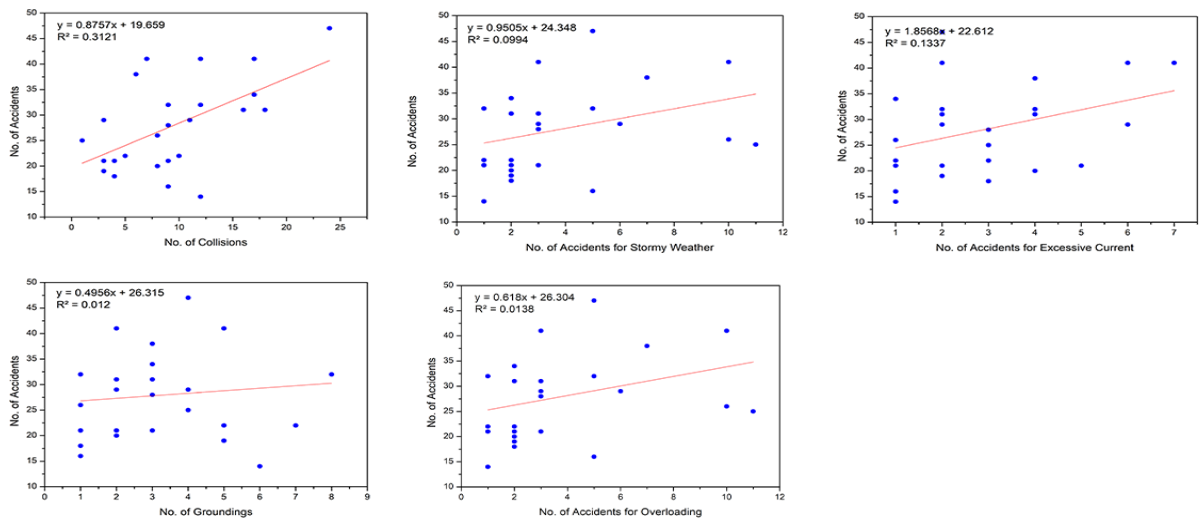

**Figure 4: Linearity check of independent variables.**

# 5 RESULTS

## 5.1 BEST FIT REGRESSION MODEL

After checking linearity stepwise regression is performed to determine the best subset. In this study, five independent variables are considered. The possible combination of one variable, two variables, three variables, four variables, five variables is $2^5$. The best subsets are sorted by checking the model selection criteria. For every model, $f_{critical}$ value is less than the calculated F statistics (f value). After completing the selection of the mathematical model residual analysis is done. After plotting residual points for each independent variable all the graphs show irregular shapes which depict that the model is ready to predict the main cause of inland waterway accidents.

**Table 3: Different mathematical models with required parameters.**

| $IWA_N = f(A_C, A_{SW}, A_{EC}, A_G, A_O)$ | $IWA_N = f(A_C, A_{SW}, A_{EC}, A_G)$ | $IWA_N = f(A_C, A_{SW}, A_{EC})$ | $IWA_N = f(A_C, A_{SW})$ | $IWA_N = f(A_C)$ | Different Mathematical Model |
|---|---|---|---|---|---|
| 0.78 | 2.86 | 6.01 | 12.371 | 19.659 | a |
| 1.096 | 1.04 | 1.03 | 1.06 | 0.87 | C |
| 1.52 | 1.62 | 1.64 | 1.45 | NA | SW |
| 0.84 | 2.19 | 2.046 | NA | NA | EC |
| 0.801 | 0.81 | NA | NA | NA | G |
| 2.10 | NA | NA | NA | NA | O |
| 5.07 | 5.13 | 5.276 | 6.323 | 7.48 | s |
| 0.74 | 0.72 | 0.688 | 0.53 | 0.31 | $R^2$ |
| 0.67 | 0.66 | 0.643 | 0.488 | 0.28 | $R^2_{adj}$ |
| 25.72 | 26.29 | 27.84 | 39.983 | 56.019 | $MSE_{(residual)}$ |
| 6.0 | 5.4 | 5.7 | 15.2 | 29.1 | $C_p$ |
| 10.76 | 12.81 | 15.43 | 12.42 | 10.434 | f |
| | | | | | $f_{critical}$ (0.05) |

# 6 DISCUSSION

In this study, five major reasons behind most of maritime accidents are considered independent variables to develop a different mathematical model. Table 3 represents the developed regression model with coefficients and the required statistical parameter considering the five independent variables. This statistical parameter needs to justify the passing criteria of the model. One thing is mentionable; this table represents the best subsets which are sorted from 25 combinations.

This research examines five primary causes of maritime accidents, considering them as independent variables in the development of a new mathematical model. Table 3 shows the resulting regression model, including coefficients and relevant statistical parameters for the five independent variables. The statistical parameter must meet the model's criteria for accuracy. Table 3 shows the most effective subsets, which have been selected from a total of $2^5$ different combinations.

It is easy to understand from table 3, the inclusion of independent variables successively, and the increasing the value of $R^2$ (multiple determination) which is the main criteria for choosing the best subset of the model simultaneously to justify the adequacy of the model. According to the highest value of adjusted $R^2$, lower value of mean square error (MSE), and minimum $C_P$ value, the best subset model is IWA (number) f $(A_C, A_{SW}, A_{EC}, A_G)$. But considering the highest value of $R^2$ with all the accidents, selected the final model for estimating the reasons behind maritime fatal are as follow:

Table 3 provides a clear illustration of how the increase in the value of $R^2$ (coefficient of multiple determination), along with the inclusion of independent variables, which is the primary criterion for choosing the best subset as well to justify the adequacy of the model. Besides, the selection criteria for this model include the highest adjusted $R^2$ value, the lowest mean square error (MSE) value, the minimum standard error value, and the minimum $C_P$ value. Based on these criteria, the best model is identified as $IWA_N$, f $(A_C, A_{SW}, A_{EC}, A_G, A_o)$ because most of the criteria are in line with the model validation requirements**(ref)**. However, by considering all independent variables (accident parameters), the final model chosen for identifying the influential parameters and estimating the number of maritime fatalities is based on the highest $R^2$ value, and the developed mathematical model is as follows:

**Best fit model: y= 0.782+ 1.096*C + 1.52*SW +0.801*G+2.1*O+ 0.84*EC………()**

Although there is no significant difference in the values of $R^2$, adjusted $R^2$, standard error, and mean square error (MSE) between the models $IWA_N$ = f $(A_C, A_{SW}, A_{EC}, A_G, A_o)$ and $IWA_N$ =f $(A_C, A_{SW}, A_{EC}, A_G)$, it is recommended to use the former model for assessing the accident because it has an additional independent variable. From the best-fit model (equation number ) it is clear that the most influential parameter ( high coefficient value 2.1) for the accident is the overloading of the ship. Practically it is also found that most of the accidents are occurred for overloading in Bangladesh (need a good reference, discuss with jubaer). The next two influential parameters are

Another thing about G (grounding), adding this $R^2$ didn't improve significantly, and also it has lower coefficient of 0.8. So, it is clear that grounding causes the lesser amount of accidents.

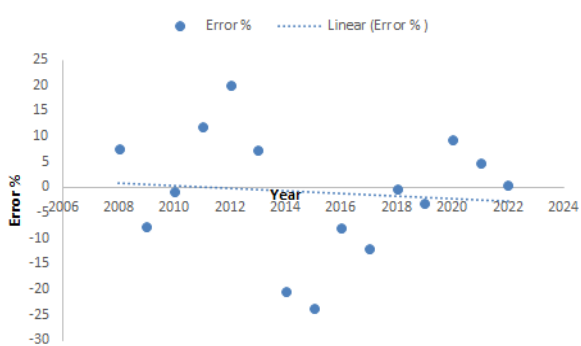
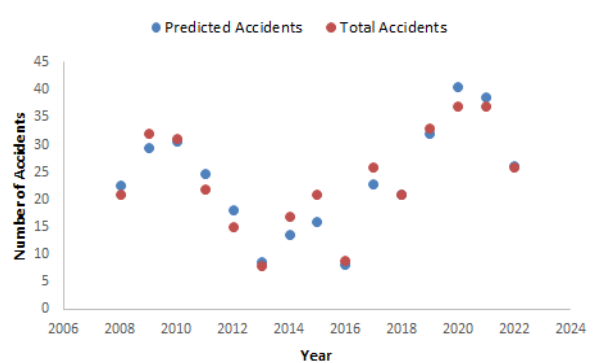

**Figure 5: Along with last 03 years comparison between predicted & total accidents using Best fit model**

We compared best fit model result with last 03 years data and found near 9.45% maximum error. Besides all theoretical explanations, something is there we need to discuss with practical point of view. SW (stormy weather) and C (collision) have a very close relationship and sometimes they are liable to each other. Collison can occur because of stormy weather. Even EC (excessive current) can be a result of SW. If we merge our model to our practical life, we may see one variable is triggering another one that is causing the accident.

Again, we get a high co-efficient of 2.10 for O (overloading). This doesn't improve the $R^2$ value as well but its high co-efficient refers that it has a huge impact on maritime accidents.

## 7 CONCLUSION AND RECOMMENDATIONS

In low-income developing countries like Bangladesh, it is natural that people choose sea as a way of travel due to low fares. And if its waterways regain its' lost heritage, pressure on road vehicles, pollution and accidents will also reduce significantly. It is not accepted to hear wails or cries of lost loved ones due to inland waterway accidents. It is everyone's hope that the waterways and its' transportations should be safe, the waterways and it's transportations should be peaceful. This research paper analyzes and picks out the most common, important independent variables which mainly dominate the total count of maritime accidents for selected locations around Bangladesh. This article makes also a possible strong relationship between dependent and independent variables in the form of a mathematical equation by using multiple linear regression. This equation is verified with the required statistical parameters to justify the adequacy of the model for the system. And applying this model, it is also found predicated result near to the practical result. The author, therefore, intends to put forward some recommendations. These recommendations may easily be implemented right now onwards in order to avoid these marine accidents.

- Foggy weather and insufficiency of navigation tools and marking are a reason behind collision-type accidents. A ship needs to control with more caution in foggy weather and a sufficient amount of navigation tools and marking need to be installed and should be checked properly before every departure.
- Sometimes maritime professionals are not enough skilled to run a ship in stormy weather or excessive current time. There is shortage of basic training as well as route information. So, more practical based training programs need to be added.
- Overloading is another reason for accidents. During holidays most of the vessels take more passengers and goods than its capacity for extra profit. Overloading disturbs the stability of a ship and sometimes it capsizes.
- The use of AI (Artificial Intelligence) should be introduced to improve the maritime safety situation of the country. The use of data science and machine learning can produce some useful assumptions about weather conditions. It can reduce accidents due to foggy or uncooperative weather.
- Mass awareness should be taken on a serious note. Most of the passengers are unaware of safety issues, they do not even know how to use life jackets, lifebuoys, etc. This is causing unwanted casualties. Live-saving appliances and safety measures should be ensured.

Further, most of the ships are not maintained properly in Bangladesh. Often minor problems are overlooked and not fixed in time. As a result, this is causing higher operation costs. So, these problems should be checked before every trip. Also, respective authority should look after vessels following Inland Shipping Ordinance rules more carefully. Most of the local vessels don't follow those rules. So, pre-analyses of vessels should be completed under close monitoring. The authority should conduct an effective investigation of every accident. The neutral investigation will help the authority to take proper decisions to prevent maritime accidents.

**ACKNOWLEDGEMENTS**

The authors would like to thank Department of Shipping (DOS), Bangladesh Inland Water Transport Authority (BIWTA) and the Accident Research Institute (ARI), Bangladesh for giving the data and other required information, which also helped immensely in this research. The authors also acknowledge the technical support from received from **** while conducting the research.